\documentclass[conference]{IEEEtran}
\usepackage{amsmath}
\usepackage{url}
\usepackage{graphicx}
\usepackage{caption}
\usepackage{subcaption}
\usepackage{multirow}
\usepackage{cite}

\begin{document}
\title{A Study of the Lexicography of Hand Gestures During Eating}

\author{\IEEEauthorblockN{Yiru Shen}
\IEEEauthorblockA{Department of Electrical and\\Computer Engineering\\
Clemson University\\
Clemson, SC\\
yirus@clemson.edu}
\and
\IEEEauthorblockN{Eric Muth}
\IEEEauthorblockA{Department of Psychology\\
Clemson University\\
Clemson, SC\\
muth@clemson.edu}
\and
\IEEEauthorblockN{Adam Hoover}
\IEEEauthorblockA{Department of Electrical and\\Computer Engineering\\
Clemson University\\
Clemson, SC\\
ahoover@clemson.edu}}

\maketitle

\begin{abstract}
This paper considers the lexicographical challenge of defining actions a person takes while eating.
The goal is to establish objective and repeatable gesture definitions based on discernible intent.  Such a standard would support the sharing of data and results between researchers working on the problem of automatic monitoring of dietary intake.
We define five gestures: taking a bite of food (bite), sipping a drink of liquid (drink), manipulating food for preparation of intake (utensiling), not moving (rest) and a non-eating category (other).
To test this lexicography, we used our definitions to label a large data set and tested for inter-rater reliability.
The data set consists of a total of 276 participants eating a single meal while wearing a watch-like device to track wrist motion.
Video was simultaneously recorded and subsequently reviewed to label gestures.
A total of 18 raters manually labeled 51,614 gestures.  Every meal was labeled by at least 1 rater, with 95 meals labeled by 2 raters. 
Inter-rater reliability was calculated in terms of agreement, boundary ambiguity, and mistakes.
Results were 92.5\% agreement (75\% exact agreement, 17.5\% boundary ambiguity).  Mistakes of intake gestures (0.6\% bite and 1.9\% drink) occur much less frequently than non-intake gestures (16.5\% utensiling and 8.7\% rest).  Similar rates were found across all 18 raters.  
Finally, a comparison of gesture segments against single index labels of bites and drinks from a previous effort showed an agreement of 95.8\% with 0.6\% ambiguity and 3.6\% mistakes.
Overall, these findings take a step towards developing a consensus lexicography of eating gestures for the research community.

\end{abstract}

\IEEEpeerreviewmaketitle

\section{Introduction}
\label{intro}
This paper considers the problem of the lexicography of defining gestures a person makes while eating.
We propose and test a vocabulary of actions to quantify gestural behaviors while eating based on discernible intent.
The set of gestures include taking a bite of food (bite), sipping a drink of liquid (drink), manipulating food for preparation of intake (utensiling), and not moving (rest).
All other activities such as using a napkin or gesturing while talking are grouped into a non-eating category (other).
We test the lexicography by labeling segments of wrist motion according to the gesture set.
The wrist motion was recorded by a watch-like device worn by participants while eating an unscripted meal \cite{Dong09}.
Synchronized video was simultaneously recorded so that the activities of the participant could be identified.
This paper describes detailed definitions of the gestures to inform human raters manually labeling the data.
It reports inter-rater reliability in order to quantify the difficulty of labeling in terms of agreement, where raters agree on the identity of activities with less than 1 second boundary disagreement, and in terms of boundary ambiguity, where the start or end of an action may be difficult to identify, and in terms of mistakes, where a rater clearly makes an error.
A total of 18 human raters labeled data for 276 subjects eating a single meal in a cafeteria setting.

A common lexicography of eating gestures is needed to support research in automated dietary monitoring.  It helps research groups compare results and share data, measure progress, and identify areas where current methods fail.  However, the lexicology in this domain is challenging due to the lack of standard definitions of terms defining actions one might take while eating.
For example, a ``bite'' may refer to the action of placing food into the mouth for consumption, but may also refer to the compressive motion of the jaw on food already in the mouth.  A ``drink'' may refer to a single instance of beverage intake, or a full container of beverage.  Several terms may be used to describe the manipulation of food prior to intake, such as cutting (to disassemble large pieces into smaller pieces), stirring (to combine foods), and dipping (to apply condiments).  Some actions do not have standard terms, such as moving food onto a fork in preparation for bringing it to the mouth.  In order to quantify eating behaviors it is necessary to establish objective, repeatable definitions.
These can then be used to label ground truth for research into automatic segmentation and classification of wrist motion to quantify eating behaviors \cite{Ramos15, Shen16}.

A lexicography of gestures can be established either top-down or bottom-up.  In the top-down approach, meanings are defined first, with gestures designed to communicate the meanings.  The most common example is sign language.  A large lexicon of both single-handed and two-handed gestures has been defined for common words and finger spelling for communication of obscure words or proper nouns \cite{Starner97, Starner98}.
Another example is vision-based interfaces in video games, where sets of gestures have been defined to characterize commands of playing a game \cite{Kang04, Wachs11}.
A third example is traffic navigation, in which different hand and body gestures have been defined \cite{Singh05, Wu10}. 
Other examples can be found in human-computer interaction (HCI), in which gestures have been defined as the commands to help user interact with the computer \cite{Pavlovic97, Rautaray15}.

In the bottom-up approach, the problem is to define gestures describing common activities that were not intended to communicate meaning \cite{Feldman91, Chen12, Lara13}.
For example, in the domain of daily activity recognition, a common lexicon includes walking, jogging and similar physical activities \cite{Mantyjarvi01, Minnen05, Parkka06, Yang07, Choudhury08, Kwapisz11}.
Independent of domains, \cite{Lausberg09} suggests some common rules for encoding hand gestures.  This paradigm is followed in this work and described more later.

\begin{figure}
\centering
  \begin{subfigure}[t]{0.45\textwidth}
    \centering
    \includegraphics[width=\linewidth]{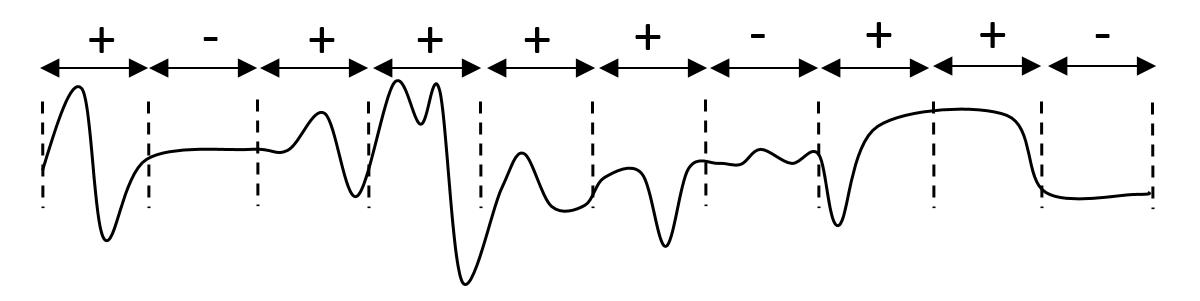} 
    \caption{Window-based. +/-: event occurs or not.}
    \label{fig:window-based}
  \end{subfigure}
  
  \begin{subfigure}[t]{0.45\textwidth}
    \centering
    \includegraphics[width=\linewidth]{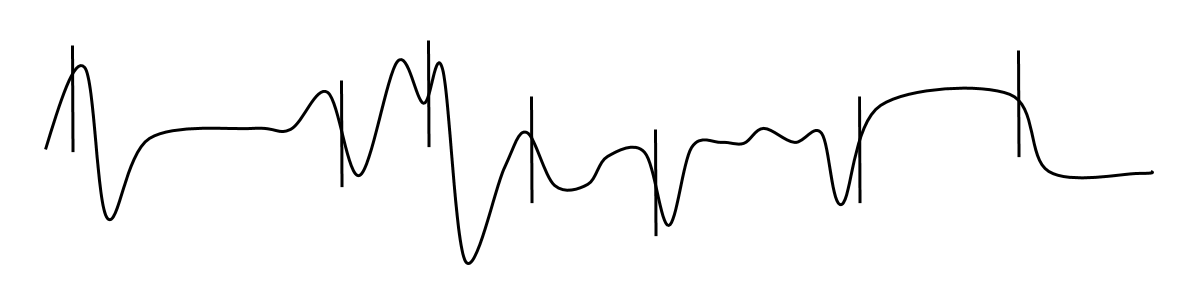} 
    \caption{Index-based. Vertical bar indicates event.}
    \label{fig:index-based}
  \end{subfigure}
  
  \begin{subfigure}[t]{0.45\textwidth}
    \centering
    \includegraphics[width=\linewidth]{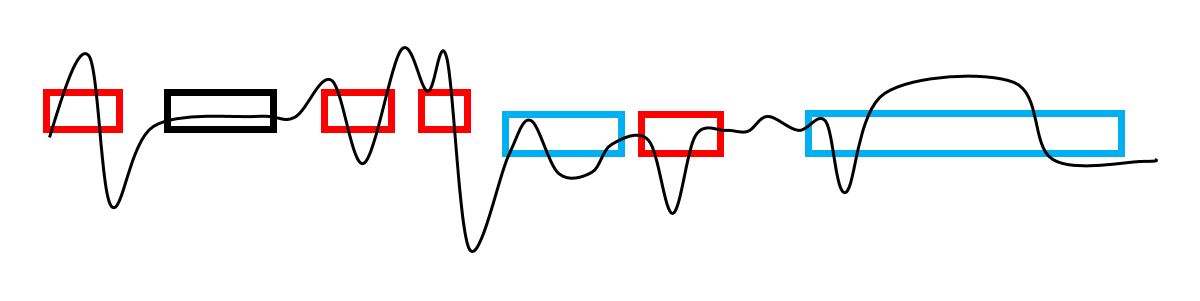} 
    \caption{Segment-based. Variable length of segments with different colors indicate different event types.}
    \label{fig:segment-based}
  \end{subfigure}
  \caption{Different approaches to annotating activity data during eating.}
  \label{fig:three-label-methods}
\end{figure}

In the domain of eating activity recognition, three approaches have been taken to label data: window-based, index-based and segment-based.  Figure \ref{fig:three-label-methods} illustrates an example of each.
In the window-based approach, equal intervals of data are labeled to indicate whether a particular type of eating event (for example, swallowing) occurs or not \cite{Sazonov08, Makeyev12, Sazonov12, Kalantarian15}.
An advantage of the window-based approach is that it simplifies the manual labeling of data.
However, it does not indicate the exact time instant or duration of the event.
In the index-based approach, a single time index is labeled when an event (for example, a bite or drink) occurs \cite{Dong12, Dong14, Passler14, Papapanagiotou17, Shen17}.
This approach can be used to more precisely identify specific times of key events, but cannot describe events that occur across a range of time, such as cutting or stirring. 
In the segment-based approach, sequences of time of variable duration are labeled, including the start and end index and the event type \cite{Amft05, Ramos15, Shen16, Zhang16}.
This approach provides more information but can be difficult to use for manual labeling as additional criteria must be determined by the human rater.
 
This paper describes a study of building a vocabulary of eating actions using segment-based labeling.
We propose a detailed definition of a set of eating gestures based on discernible intent.
The definitions were written with the goal of being objective with as little ambiguity as possible so that raters can follow the same strategy to label eating activities.
We tested the definitions on a large data set of 276 participants eating in a cafeteria setting and report several inter-rater reliability tests.

The remainder of this paper is organized as follows. 
Section \ref{sec:methods} describes the experimental conditions, gesture definitions and inter-rater reliability including agreement, boundary ambiguity and mistake to access the definition reliability, followed by the comparison between intake gestures and index-based labels.
Section \ref{sec:results} presents a set of results and explains their implications.
Section \ref{sec:discussion} concludes the paper and discusses the future work.

\section{Methods}
\label{sec:methods}
\subsection{Data}
\label{subsec:data}

\begin{figure}
\begin{center}
\begin{tabular}{c c}
\includegraphics[height=1.3in]{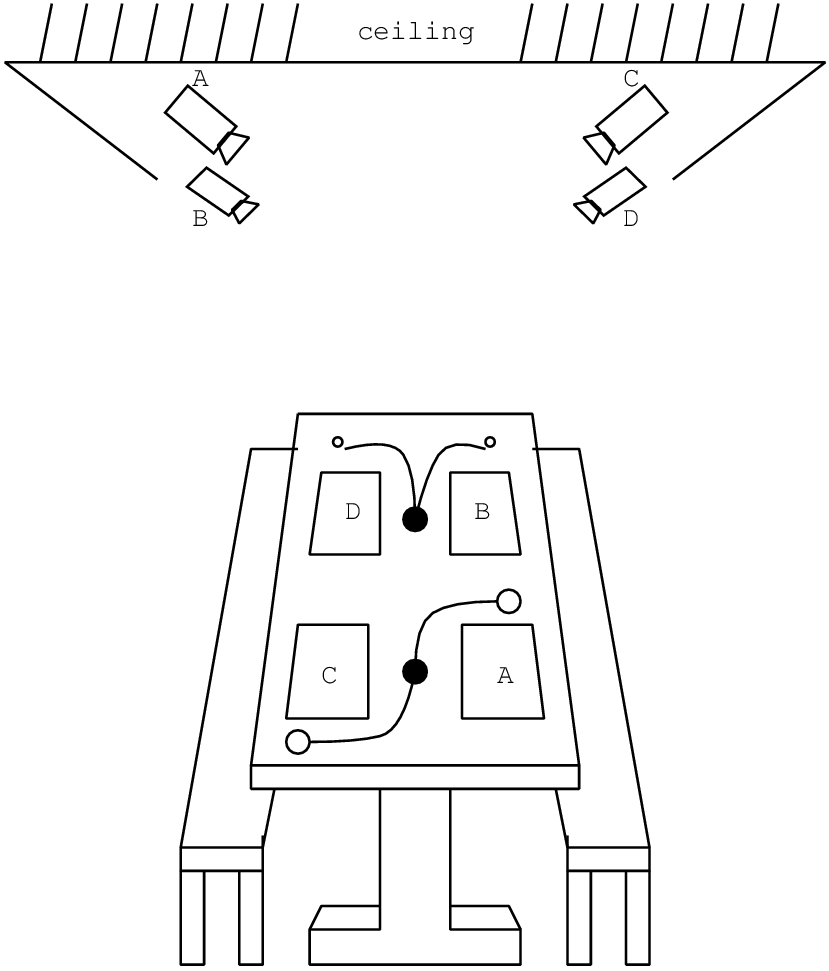} &
\includegraphics[height=1.3in]{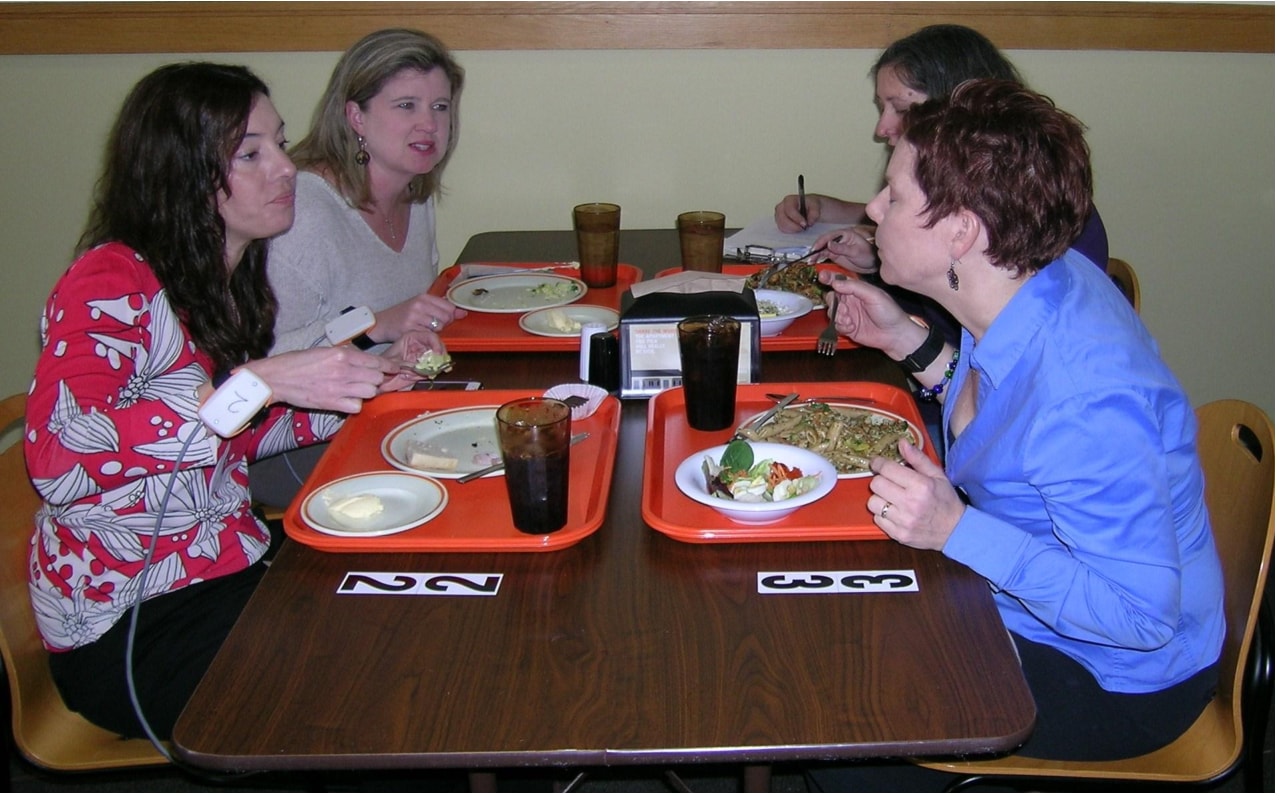} 
\end{tabular}
\end{center}
\caption{The table instrumented for data collection.
Each participant wore a custom tethered device to track wrist motion.}
\label{fig:eating_table}
\end{figure}

Data recording took place in the Harcombe Dining Hall at Clemson University.
Figure \ref{fig:eating_table} shows an illustration of an instrumented table that can record data up to four participants simultaneously \cite{Huang13}. 
Four digital video cameras in the ceiling (approximately 5 meters height) were used to record each participant's mouth, torso, and tray during meal consumption.
A custom device was designed to record the wrist motion of subjects at 15 Hz during eating, using MEMS accelerometers (STMicroelectronics LIS344ALH) to measure acceleration of x, y and z axis, and gyroscopes (STMicroelectronics LPR410AL) to measure rotational velocity around yaw, pitch and roll. 
Cameras and wrist motion trackers were wired to the same computers and used timestamps for synchronization.

A total of 276 participants were recruited and each consumed a single meal \cite{Shen16}.
Participants were free to choose any foods and beverages available. 
Upon sitting at the table to eat, an experimental assistant placed the wrist motion tracking device on the dominant hand of the participant and interviewed them to record the identities of foods selected.
The participant was then free to eat naturally. 
For 5 participants, either the video or wrist motion tracking failed to record, and for 2 participants non-dominant hands were used for recording; these are excluded from analysis.

\begin{figure*}
	\centering
	\includegraphics[width=\textwidth, height=9.7cm]{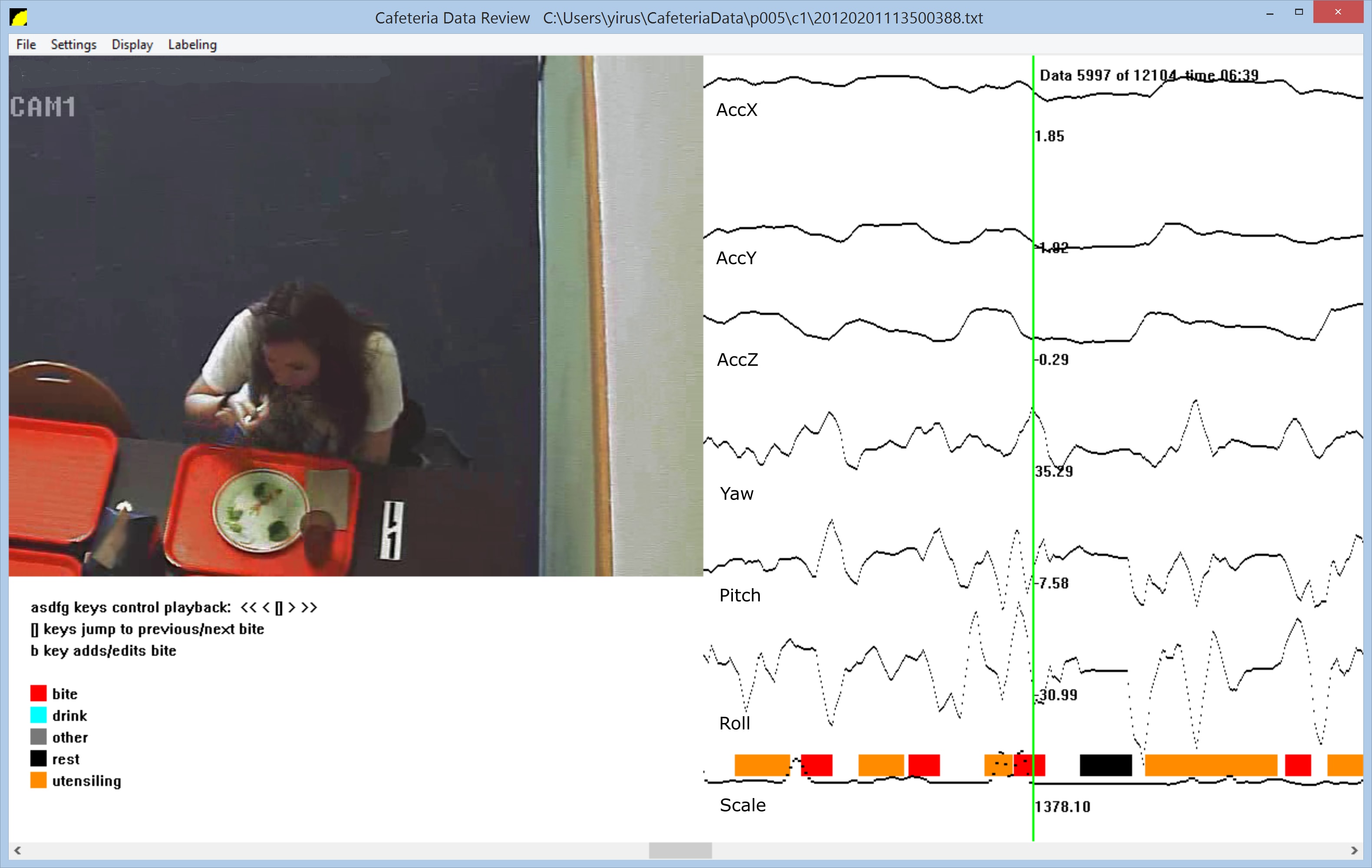}
    \caption{A custom program for gesture labeling. Box with different colors indicate gesture types: red = bite, aqua = drink, orange = utensiling, black = rest and grey = other. }
	\label{fig:label-tool}
\end{figure*}

\subsection{Definitions of Gestures}
\label{subsec:definition}

Our proposed lexicon was motivated by separating intake related gestures from non-intake related gestures.  There are arguably fewer of the former compared to the latter.  Since the primary goal in research into automated monitoring of dietary intake is to quantify intake, our proposed lexicon uses a relatively small set of non-intake gestures.

We define four eating-related gestures (two intake and two non-intake): \textit{bite}, \textit{utensiling}, \textit{drink}, \textit{rest}.
All other activities (e.g. gesturing while talking, cleaning with a napkin etc.) are referred to as a fifth gesture \textit{other}. 
Following the paradigm proposed in \cite{Lausberg09}, each gesture is defined consisting of the following parts with at least 1 second duration:

\renewcommand{\labelenumi}{(\alph{enumi})}
\begin{enumerate}
\item the description of the activity;
\item the start time of the activity;
\item the end time of the activity;
\item particular events that should be included or excluded;
\end{enumerate}

\noindent
\textbf{Bite}
\begin{enumerate}
\item The subject puts food into their mouth.
\item Starts when a hand or utensil starts moving towards the mouth.
\item Ends when the hand or utensil finishes moving away from the mouth.
\item Bites need not begin and end at a plate. Motion towards and away from the mouth should define the boundaries; with food consumption taking place in between.
\item A single bite may include multiple successive back-and-forth motions from a utensil or hand to the mouth, that individually did not complete the hand motion away from the mouth, and that were separated by less than 1 second of time.
\end{enumerate}

\noindent
\textbf{Drink}
\begin{enumerate}
\item The subject puts beverage into their mouth.
\item Starts when a hand begins moving a beverage towards the mouth.
\item Ends when the hand has finished moving away from the mouth.
\item Each individual sip should be a different drink (if multiple sips are taken).
\end{enumerate}

\noindent
\textbf{Utensiling}
\begin{enumerate}
\item The subjects uses an utensil or their hand(s) to manipulate, stir, mix or prepare food(s) for consumption.
\item Starts when manipulating the food.
\item Ends when manipulating has finished.
\item This includes moving food around the plate, dipping foods in sauces, cutting foods, and other similar activities. 
\end{enumerate}

\noindent
\textbf{Rest}
\begin{enumerate}
\item The subject's dominant hand has little or no motion. The range of motion that may be considered rest depends upon the individual. Different people have different levels of physiological tremor (motion that occurs in everyone and has no medical significance) and thus the threshold for maximum motion during rest will vary subject to subject.
\item The determination of rest should be based on the instrumented dominant hand only.
\item Starts when there is no intent (subject's hand stop moving).
\item Ends when new intent becomes apparent (subject's hand begins moving again with clear intent for at least 1 second).
\item A period of rest may include time when a person is holding a utensil, food or drink, but where the instrumented hand is relatively motionless. 
\end{enumerate}

\noindent
\textbf{Other}
\begin{enumerate}
\item All other actions should be left unlabeled. Examples include reaching towards food (e.g. prior to a bite gesture), gesturing while talking, cleaning with a napkin, and moving a plate.
\item In cases where the action of the instrumented hand alone is unclear, the subject's face and body can be viewed to help discern intent. For example, if the subject is talking and there is a slight motion in the instrumented hand, one may assume it is gesturing while talking (other) instead of rest.
\item In cases where it is difficult to differentiate between rest and other, or utensiling and other, the other label is preferred.  
\end{enumerate}

\subsection{Inter-rater Reliability}
\label{subsec:IRR-gestures}

In order to test  inter-rater reliability we developed methods to compare multiple labelings of the same meal recording.
The data provided by raters is also intended to be used in the future for training classifiers and evaluating automatic segmentation, and thus the process for combining multiple raters' labels into a union is described.
Since our data set is so large and was collected during natural (unscripted) eating, the total process took more than 700 man-hours of work.

Figure \ref{fig:label-tool} shows a custom program we built to facilitate labeling.
The tool was coded using Microsoft Visual Studio. 
The left panel displays the video while the right panel shows the synchronized wrist motion tracking data. 
Top to bottom on the right panel shows the 6 axes of motion (AccX, AccY, AccZ, yaw, pitch and roll) with a seventh line at the bottom indicating tray weight as measured from a table embedded scale.
Keyboard controls allow for play, pause, rewind and fast forward.
Vertical green line indicates the time currently displayed in the video.
A human rater annotates a meal by watching the video and uses frame-by-frame rewinding and forwarding to identify the start and end time and type of a gesture according to our definitions.  
Boxes laid over the seventh line indicate periods of time labeled as gestures (for example, red = bite). 
Unlabeled segments with duration longer than 4 seconds are considered as type other, unlabeled segments shorter than 4 seconds are considered transitions between gestures and are ignored \cite{Ramos15}.
The process of labeling a single meal took 60-120 minutes.

Due to the work-intensive nature of this labeling process, only 95 meals (20\%) were labeled by two raters.
In total 18 raters contributed to the process.  
Raters were trained in several training sessions to understand the process and the definitions of gestures.

\begin{figure*}
\centering
  \begin{subfigure}[t]{0.15\textwidth}
    \centering
    \includegraphics[width=\linewidth]{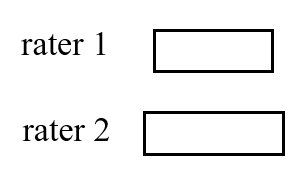} 
    \caption{Agreement.} \label{fig:taba2}
  \end{subfigure}%
  \begin{subfigure}[t]{0.15\textwidth}
    \centering
    \includegraphics[width=\linewidth]{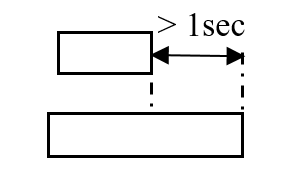} 
    \caption{BA I.}\label{fig:taba4}
  \end{subfigure}%
  \begin{subfigure}[t]{0.15\textwidth}
    \centering
    \includegraphics[width=\linewidth]{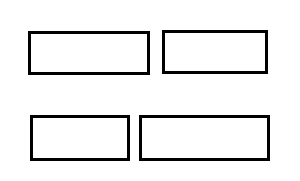} 
    \caption{BA II.}  \label{fig:taba5}
  \end{subfigure}%
  \begin{subfigure}[t]{0.15\textwidth}
    \centering
    \includegraphics[width=\linewidth]{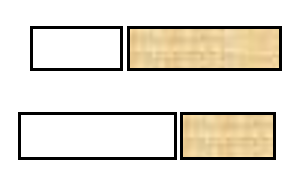} 
    \caption{BA III.}  \label{fig:taba5}
  \end{subfigure}%
  \begin{subfigure}[t]{0.15\textwidth}
    \centering
    \includegraphics[width=\linewidth]{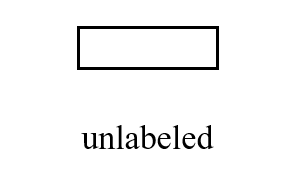} 
    \caption{Mistake-missed.} \label{fig:taba3}
  \end{subfigure}%
  \begin{subfigure}[t]{0.15\textwidth}
    \centering
    \includegraphics[width=\linewidth]{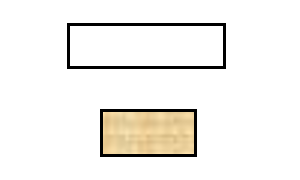} 
    \caption{Mistake-identity.}  \label{fig:taba5}
  \end{subfigure}
\caption{Different cases of gesture matching between two raters. Segments with different colors represent different identities. BA: boundary ambiguity. }
\label{fig:IRR-segments}
\end{figure*}

Quantifying rater agreement is complicated because labeling contains multiple steps. 
First, each rater had to decide the start and end index of a gesture.
Second, they had to identity the type of a gesture.
Therefore we developed a custom approach that includes the following 6 cases to determining gesture matching as illustrated in Figure \ref{fig:IRR-segments}.
For each gesture labeled by one rater, any overlapped gesture labeled from the second rater was examined. 

\noindent
\textbf{Agreement.} 

\noindent
If only one corresponding gesture with the same identity was matched and the disagreement of both start and end index is within 1 second, then the start and end time index were taken as the averaged time indicated by two raters. 

\noindent
\textbf{Boundary Ambiguity I.} 

\noindent
If only one corresponding gesture with the same identity was matched and the disagreement of start or/and end index is longer than 1 second, and there are no other gestures labeled within the boundaries, then the start and end index were taken as in Equation \ref{eq:index}.
\begin{equation}
t = \begin{cases} 
	(t_{1}+t_{2})/2, & \text{$||t_{1}-t_{2}||\leq \mbox{1 sec}$} \\
	max(t_{1}, t_{2}), & \text{otherwise}
\end{cases}
\label{eq:index}
\end{equation}
where $t_{1}$ and $t_{2}$ represent index labeled by rater \#1 and rater \#2, and $max$ indicates the index providing the maximum gesture extent.
For intake gestures, this is usually caused by a pause at the start or end of a gesture, e.g. during taking a bite the participant did not complete moving food towards the mouth until a pause for masticating food from a previous bite. 
For non-intake gestures, this is usually caused by some unintentional ambiguity in the definitions. 
For example, when dipping food in a sauce, one rater may label the motion of reaching towards the sauce as utensiling while a second rater starts the utensiling when the food touches the sauce.

\noindent
\textbf{Boundary Ambiguity II.} 

\noindent
If multiple corresponding gestures with the same identity were matched, two cases are discussed here. 
For intake gestures, a corresponding match with single index-based labels created in a previous effort \cite{Shen16} were compared.
Gestures from the rater which matched the index-based labels in terms of the amount of intake events were considered correct, and the longer duration of the two raters were taken as the union.
For non-intake gestures, the max extent of the start and end index of the gesture was taken. 
Boundary Ambiguity II (N:N or N:1 matching) is usually caused by one rater labeling a single whole period of time as a single gesture while another rater segmented it into multiple sub-periods.

\noindent
\textbf{Boundary Ambiguity III.} 

\noindent
If only one corresponding gesture with the same identity was matched and the disagreement of start or/and end index is longer than 1 second, and there is another gesture labeled within the boundaries but it was matched against a different gesture, then the start and end index were taken as in Equation \ref{eq:index}. 
If the extra gesture did not match anything, then the gesture in query is considered as matched but the extra gesture is considered as mistake-identity.

\noindent
\textbf{Mistake-missed.} 

\noindent
If no corresponding gesture was matched, then the gesture in query was taken as the union. 
For intake gestures, this is usually caused by one rater missing an action. 
For non-intake gestures, this is usually caused by the ambiguity of definitions. 
For example, one rater may label rest for a participant while another rater may consider the same period of time as a gap.

\noindent
\textbf{Mistake-identity.} 

\noindent
This happens when one corresponding gesture with a different identity was found.
This is usually caused by a rater incorrectly identifying a gesture.

\begin{figure*}
\centering
\includegraphics[width=\linewidth, height=3cm]{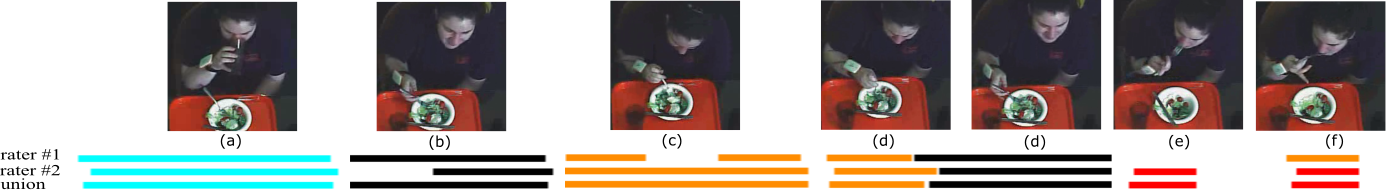} 
\caption{Example of gesture matching. From top to bottom: gestures labeled by rater \#1, rater \#2 and the union. (a)-(f) illustrate different cases of gesture matching. Red = bite, aqua = drink, orange = utensiling, black = rest.}
\label{fig:GT-gestures}
\end{figure*}

Using this process, rater performance can be evaluated using three metrics: agreement, boundary ambiguity (I, II, and III), and mistake (mistake-missed and mistake-identity).
Figure \ref{fig:GT-gestures} shows an example of gesture matching and the union labels. 

\subsection{Comparing Intake Gestures with Index-based Labels}

In a previous work our group labeled this same data set with single time indices indicating when bites and drinks occurred \cite{Shen16}.  The time indices indicated when the food or beverage first touched the mouth initiating intake.  We used this set of time indices to further test the segment labels. Note that this could only be done for intake gestures as events corresponding to utensiling, rest and other were not labeled in the previous work.

Each intake gesture (bite and drink) was searched for any corresponding index-based labels.
If one gesture contained exactly one index-based label within its boundaries, then it was considered as agreement.
If one gesture contained multiple index-based labels within its boundaries, then it was considered as ambiguity.
This usually happened when a rater labeled one long segment that contained multiple short bites or drinks.
If no index-based label was contained within a gesture, or an index-based label was not matched against any gesture, then it was considered as missed.
Note that the index-based labels searched for matching intake gestures were only considered if they were performed by the dominant hands of the participants.

\section{Results.}
\label{sec:results}

Table \ref{table:statitics-gestures} lists the distribution statistics of the five gestures labeled by all raters.
The minimum durations of bite, drink, rest and utensiling were enforced to be 1 second; the minimum duration of other was enforced to be 4 seconds by definition.
Notice that utensiling and rest could last up to 3 and 6 minutes, respectively.
For utensiling, this happened when a subject took a long time peeling a tangerine.
For rest, this happened when a subject talked to other people for a long time while the hand wearing device was at rest.
The variable duration of gestures demonstrates the challenge of labeling segments: it is difficult to accurately label the start and end index. 

\begin{table}
\centering
\begin{tabular}{| c | c | c | c | c |} \hline 
Type & \#Gestures & \multicolumn{3}{c|}{Duration (sec)} \\  \cline{3-5}
&  &Average $\pm$ Stddev & Min & Max\\ \hline
Bite & 18462 & 2$\pm$1 & 1 & 11 \\ \hline
Drink & 2182 & 6$\pm$2 & 1 & 18 \\ \hline
Utensiling & 14861 & 5$\pm$5 & 1 & 186 \\ \hline
Rest & 14761 & 8$\pm$12 & 1 & 341 \\ \hline
Other & 1348 & 9$\pm$6  & 4 & 73 \\ \hline
\end{tabular}
\caption{Statistics of gestures.}
\label{table:statitics-gestures}
\end{table}

Table \ref{table:2-raters-match} lists the inter-rater reliability for meals labeled by two raters. 
Note that only four gestures (bite, drink, rest and utensiling) are evaluated since gesture ``other'' will be automatically determined if the gap between gestures are longer than 4 seconds. 
The overall agreement is 92.5\% with exact agreement of 75\% and 17.5\% of boundary ambiguity.
The agreement for bite and drink is 99.4\% and 98.1\%, respectively.
This indicates a high degree of agreement between raters on intake related gestures.
The overall mistake rate is 7.5\% with most of the mistakes from non-intake gestures.
This is due to the nature of ambiguity on non-intake gestures. 
For example, raters may have different understanding on levels of physiological tremor in the definition of rest and one rater labeled a small amount of motion as rest while another rater did not label it. 

The usefulness of a third rater independently labeling each meal and then comparing it to the union from two raters was explored. 
After 7 meals were labeled, the process was stopped.
In those 7 meals, the mistake and ambiguity rate did not change.

\begin{table}
\centering
\begin{tabular}{|c | c | c | c | c |c |} \hline
Cases & \#Gestures & Bite & Drink & Rest & Utensiling \\ 
& (\%) & (\%) & (\%) & (\%) & (\%) \\ \hline
Agreement & 13184 & 5814 & 594 & 2926 & 3850 \\ 
& (75.0\%) & (89.6\%) & (66.9\%) & (58.9\%) & (73.5\%) \\ \hline
BA  I & 1030 & 136 & 160 & 548 & 186 \\ 
& (5.9\%) & (2.1\%) & (18.0\%) & (11.1\%) & (3.6\%) \\ \hline
BA II & 784 & 21 & 73 & 443 & 247 \\ 
& (4.5\%) & (0.3\%) & (8.2\%) & (8.9\%) & (4.7\%) \\ \hline
BA III & 1250 & 478 & 44 & 232 & 496 \\ 
& (7.1\%) & (7.4\%) & (5.0\%) & (4.7\%) & (9.5\%) \\ \hline
Mistake & 1079 & 23 & 9 & 700 & 347 \\ 
-missed & (6.1\%) & (0.4\%) & (1.0\%) & (14.1\%) & (6.6\%) \\ \hline
Mistake & 249 & 14 & 8 & 117 & 110 \\ 
-identity & (1.4\%) & (0.2\%) & (0.9\%) & (2.4\%) & (2.1\%) \\ \hline
Overall & 1328 & 37 & 17 & 817 & 457 \\ 
mistake & (7.5\%) & (0.6\%) & (1.9\%) & (16.5\%) & (8.7\%) \\ \hline
Overall  & 3064 & 635 & 277 & 1223 & 929 \\ 
BA & (17.5\%) & (9.8\%) & (31.2\%) & (24.6\%) & (17.7\%) \\ \hline
Overall & 16248 & 6449 & 871 & 4149 & 4779 \\
agreement & (92.4\%) & (99.4\%) & (98.1\%) & (83.5\%) & (91.3\%) \\ \hline
\#Gestures & 17576 & 6486 & 888 & 4966 & 5236 \\ \hline
\end{tabular}
\caption{Inter-rater reliability for meals with two raters. BA: boundary ambiguity. }
\label{table:2-raters-match}
\end{table}

Table \ref{table:match-bite-database} lists the inter-rater reliability of comparing intake gestures with index-based labels.
It can be seen that the agreement and missed rate were both improved by 1.4\% when the second rater contributed to labeling. 
The agreement and mistake rate when a third rater contributed did not change compared to gestures labeled by two raters. 
The small amount of improvement of multiple raters illustrates that a single labeling is sufficient for use in classifier development. 

\begin{table}
\centering
\begin{tabular} {| c | c | c |} \hline
 & One rater & Two raters \\ \hline
Agreement & 16029 & 3461  \\ 
(\%) & (94.4\%) & (95.8\%) \\ \hline
Ambiguity  & 93 & 21 \\ 
(\%)  & (0.5\%) & (0.6\%) \\ \hline
Missed & 862 & 128  \\ 
(\%) & (5\%) & (3.6\%) \\ \hline
\# Gestures & 16984 & 3610  \\ \hline
\end{tabular}
\caption{Inter-rater reliability between intake gestures and index-based labels in bite database \cite{Shen16}.}
\label{table:match-bite-database}
\end{table}

Table \ref{table:IRR-raters} lists the inter-rater reliability of raters who labeled at least 8 meals.
Overall the total agreements range from 89\% to 98\%.
It should be noted that even for rater YS who labeled a large amount of gestures, the total agreement had 92\% indicating a high degree of consistency for gesture definitions across the large data set. 

\begin{table}
\centering
\begin{tabular}{| c | c | c | c | c | c |} \hline
Rater & \#Gestures & Total & Agreement & BA & Mistake \\ 
&  & agreement (\%) &(\%) & (\%) & (\%) \\\hline
YS & 8481 & 7821 & 6363 & 1458  & 660  \\ 
&  & (92\%) & (75\%)  &(17\%) & (8\%) \\ \hline
RB & 942 & 864 & 743 & 121 & 78 \\ 
&  & (92\%) & (79\%)  &(13\%) & (8\%) \\ \hline
AS & 1107 & 983 & 783& 200 & 124 \\ 
& & (89\%) & (71\%)  &(18\%) & (11\%) \\ \hline
JW & 1182 & 1077 & 937 & 140 & 105 \\ 
& & (91\%) & (79\%)  &(12\%) & (9\%) \\ \hline
JP & 818 & 800 & 656 & 144 & 18 \\ 
& & (98\%) & (80\%) & (18\%) & (2\%) \\ \hline
PJ & 649 & 619 & 470 & 149  & 30  \\
& & (95\%) & (72\%)  &(23\%) & (5\%) \\ \hline
TH & 699 & 649 & 499 & 150 & 50 \\ 
& & (93\%) & (71\%)  &(21\%) & (7\%) \\ \hline
JD & 1099 & 1012 & 875 & 137 & 87 \\ 
& & (92\%) & (80\%)  &(12\%) & (8\%) \\ \hline
\end{tabular}
\caption{Inter-rater reliability for raters labeling at least 8 meals. BA: boundary ambiguity. }
\label{table:IRR-raters}
\end{table}

\section{Discussion}
\label{sec:discussion}

This paper considers the problem of the lexicography of hand gestures during eating.
Compared to the lexicography of hand gestures in sign language where the gesture vocabulary is designed top-down, the lexicography of hand gestures during eating must be designed bottom-up to encode a large variety of existing natural gesture behaviors. 
The goal of this study was to establish objective and repeatable definitions based on discernible intent during eating.
A set of vocabulary of eating actions was built to quantify gestural behaviors.
A total of 51,614 gestures were manually labeled and definitions were tested in a large data set for 276 participants.
Duration of gestures varying from 1 second to 341 seconds indicates the difficulty of labeling segments.
Inter-rater reliability of 18 raters showed 92.5\% total agreement (75\% exact agreement and 17.5\% boundary ambiguity).
The intake gestures had total agreement of 99.4\% and 98.1\% for bite and drink, respectively.
Inter-rater reliability was further tested against a previously labeled data set of single time index-based labels.  This test showed agreement of 94.4\% and 95.8\% for meals labeled by one and two raters, respectively.
The performance of raters who labeled at least 8 meals was assessed, with total agreement ranging from 89\% to 98\%.
Overall these findings show that the definitions are consistent and repeatable across a large data set.

Although the overall mistake rate is 7.5\%, most mistakes are from non-intake gestures.
By design, a large variety of patterns resides in our existing definitions of non-intake gestures.
For utensiling, stirring food and cutting food contain different patterns, where stirring involves more rotational motions while cutting involves periodic horizontal motions.
Other actions such as peeling a fruit or vegetable or mixing food are also typical in utensiling. 
For rest, people have different levels of physiological tremor which is the natural variation in capability of holding perfectly still.
Therefore our limited set of gesture labels has some difficulty in labeling all natural behaviors.
Potential future work could explore an extension of our vocabulary to include additional gesture types or subdivide some gestures into multiple types.  However, the purpose of this lexicography is to support research into the automated monitoring of dietary intake, where emphasis is on the detection and quantification of intake events.  Recognizing a wider body of non-intake events may not be helpful towards this goal.  Labeling a wider body of gesture types may also reduce inter-rater reliability.

A limitation of this work is that it was only tested on meals eaten in a cafeteria setting.  It is possible that eating related gestures in other environments may require modifications to this encoding scheme.  However, mitigating that problem is the fact that a relatively large number of people (276) were recorded eating a completely unscripted meal.  Another limitation of this work is that it only considers gestures related to wrist motion.  Other works \cite{Sazonov08, Sazonov10, Sazonov12, Makeyev12, Passler14, Kalantarian15, Papapanagiotou17} focus on eating actions related to the head or throat such as chewing and swallowing.  This work could be extended to include lexicography for those types of events and should provide some background to assist with its development.

Previous work in our group studied the temporal dependency between eating activities using hidden
Markov models and contextual information to improve recognition accuracy \cite{Ramos15, Shen16}. 
Future work will use the labeled segments as the ground truth to further investigate the sequential dependencies between gestures, in the application of classification and automatic segmentation of motion data \cite{Lee99, Yang07, Kim07}.

\section*{Acknowledgment}
We gratefully acknowledge the support of the NIH via grant 1R01HL118181-01A1.
We also wish to thank the 18 volunteers who manually labeled the gesture database for their hundreds of hours of work.

{99}

\end{document}